\documentclass[12pt]{iopart}
\usepackage{amssymb}
\usepackage{graphicx}
\usepackage{subfigure}
\usepackage{hyperref}
\usepackage{iopams}

\begin{document}

\title{General Properties of a System of $S$ Species Competing Pairwise}
\author{R. K. P. Zia}
\address{Department of Physics, Virginia Tech, Blacksburg, VA 24061, USA}
\date{December 29, 2010}

\begin{abstract}
We consider a system of $N$ individuals consisting of $S$ 
species that interact pairwise: $x_m+x_\ell \rightarrow 2x_m\,\,$
with arbitrary probabilities $p_m^\ell $. With no spatial structure, 
the master equation yields a simple set of rate equations in a mean field 
approximation, the focus of this note. Generalizing recent findings of 
cyclically competing three- and four-species models, we cast these equations
in an appealingly simple form. As a result, many general properties of such 
systems are readily discovered, e.g., the major difference between even 
and odd $S$ cases. Further, we find the criteria for the existence of 
(subspaces of) fixed points and collective variables which evolve trivially 
(exponentially or invariant). These apparently distinct aspects can be 
traced to the null space associated with the interaction matrix, $p_m^\ell $. 
Related to the left- and right- zero-eigenvectors, these appear to be 
``dual'' facets of the dynamics. We also remark on how the standard 
Lotka-Volterra equations (which include birth/death terms) can be regarded 
as a special limit of a pairwise interacting system.
\end{abstract}

\emph{Introduction.} Population dynamics is a venerable subject, dating back
two centuries to Malthus, Verhulst, Lotka, Volterra, and many others\cite
{HS98,Now06,Frey09}. Nonetheless, new and interesting phenomena are continually
being discovered. For example, many studies of cyclic competition between 3
species (with no spatial structure, e.g., a well-mixed system) attracted
considerable recent attention \cite{BRSF09}. In fact, in 2009, Science Daily
popularized this topic\cite{SD09} by branding it ``Survival of the
Weakest.'' We extended this investigation to a system with 4 species\cite
{4SS}, which displayed no such counter-intuitive behavior. Instead, we found
an intuitively understandable principle which underpins all systems with
cyclically competing species, namely, ``The prey of the prey of the weakest
is least likely to survive.'' In the case of cyclically competing 3 species,
the prey of one's prey is also one's predator. Thus, its demise is indeed
``good news'' for the weakest and leads to the eye-catching headline. In this
short note, we considered a wider range of systems of $S$ species
interacting pairwise, with arbitrary rates. Focusing only on a mean field
description (i.e., rate equations), we find remarkable general properties,
such as fixed points, invariant manifolds, collective variables with simple
time dependence, as well as a (possibly new) form of ``duality.'' We begin
by specifying the individual based stochastic model, from which our MF
approximation is derived. This note is devoted only to the properties of
solutions to the MF equations, however.

\emph{Individual based model. }Consider a system with $N$ individuals, each
being a member of one of $S$ species. Let us denote the species by $x_m$,
with $m=1,...,S$, and the number of individuals of each by $N_m$. We allow
only pairwise interactions, i.e., 
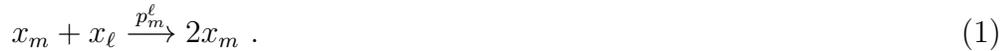
\begin{equation}
x_m+x_\ell \stackrel{p_m^\ell }{\longrightarrow }2x_m\,\,.  \label{pair}
\end{equation}
where $p_m^\ell $ are arbitrary probabilities for a ``predator'' $x_m$ to
consume a ``prey'' $x_\ell $. Note that, if we wish to model bi-directional
interactions, then these $p$'s represent the net consumption of the dominant
species. Thus, there are at most $S\left( S-1\right) /2$ such \emph{positive 
}quantities. When two individuals encounter, the role of each is well
defined: one is the predator, the other is the prey. To emphasize, we
illustrate with $p_1^5=0.7$, and both $x_1+x_5$ and $x_5+x_1$ becomes $2x_1$
with probability $0.7$ (and unchanged with probability $0.3$). Defining an
update of our system as randomly choosing a pair and letting them interact,
we see that the $N_m$'s change by $\pm 1$ or $0$ with $N=\sum_mN_m$
remaining a constant for all times. Depending on the $p_m^\ell $'s, there
are at least $S$ absorbing states.

The appropriate quantity for describing this stochastic evolution is, of
course, $P\left( \left\{ N_m\right\} ;\tau \right) $, the probability to
find the system with the set $\left\{ N_m\right\} $ after updating it $\tau $
times from some given initial $P\left( \left\{ N_m\right\} ;0\right) $. The
change in one step, $P\left( \left\{ N_m\right\} ;\tau +1\right) -P\left(
\left\{ N_m\right\} ;\tau \right) $, is given by $2/N\left( N-1\right) $ 
times
\begin{equation}
\hspace{-1.0cm} \sum_{\left\langle n,\ell \right\rangle
}p_n^\ell \left[ \left( N_n-1\right) \left( N_\ell +1\right) P\left( \left\{
N_m-\delta _{mn}+\delta _{m\ell }\right\} ;\tau \right) -N_nN_\ell P\left(
\left\{ N_m\right\} ;\tau \right) \right]  \label{ME}
\end{equation}
where the sum is over only predator-prey pairs. Finally, to include the
standard form of a Lotka-Volterra model, we should add birth/death
probabilities 
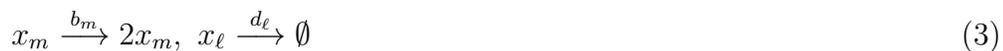
\begin{equation}
x_m\stackrel{b_m}{\longrightarrow }2x_m,\,\,x_\ell 
\stackrel{d_\ell }{\longrightarrow }\emptyset  \label{bd}
\end{equation}
as well. Except for the last section, in which we will comment on this
addition, we consider only pair interactions in the remainder of this note.

\emph{Rate equations and general properties of their solutions.} 
To proceed, we exploit the standard MF approximation.
Taking the large $N$ limit, rescaling time by $t\equiv \tau /N$ (which
becomes a continuous variable), and replacing averages 
\[
\mathcal{O}\left( \tau \right) \equiv \left\langle \mathcal{O}\right\rangle
_\tau \equiv \sum_{\left\{ N_m\right\} }\mathcal{O}\left( \left\{
N_m\right\} \right) P\left( \left\{ N_m\right\} ;\tau \right) 
\]
of products of $N_m$ by the product of the averages, we arrive at the rate
equations for the fractions 
\begin{equation}
\left\langle X_m\right\rangle _\tau \equiv \left\langle N_m\right\rangle
_\tau /N  \label{X}
\end{equation}
namely, 
\begin{equation}
\partial X_m\equiv \frac{dX_m}{dt}=X_m\sum_\ell k_{m\ell }X_\ell \,.
\label{X-eqn}
\end{equation}
Here, 
\[
k_{m\ell }=-k_{\ell m}=2p_m^\ell 
\]
can be regarded as the elements of an antisymmetric matrix $\mathbb{K}$.
Although eqn. (\ref{X-eqn}) is nonlinear, its special form allows us to
write a quasi-linear equation
\footnote{%
A special system, with a form similar to (\ref{X-eqn}), can be solved
exactly \cite{Now06}. Instead of a matrix, $k_{m\ell }$, there is just a
constant vector $g_\ell $. Such a system is suitable for species competiting
for the same pool of resources, which gets depleted via the combination $%
\sum_\ell g_\ell X_\ell $. I thank Zoltan Toroczkai for pointing this case
out to me.}. 
Using vector notation, it is 
\begin{equation}
\partial \left| U\right\rangle =\mathbb{K}\left| X\right\rangle  \label{U eqn}
\end{equation}
where the elements of $\left| U\right\rangle $ are 
\begin{equation}
U_m=\ln X_m  \label{U def}
\end{equation}
The rest of this note is devoted to the consequences of eqns. (\ref{X-eqn})
or (\ref{U eqn}), apart from the obvious, strict conservation law: $%
\sum_\ell X_\ell =1$. Denoting a row vector with all its elements being
unity by $\left\langle 1\right| $, this constraint reads 
\begin{equation}
\left. \left\langle 1\right| X\right\rangle =1 \thinspace \thinspace .
\label{norm}
\end{equation}
For physical systems, we also demand $X_\ell \geq 0$, so that the
evolution of interest here takes place in an $S-1$ simplex, i.e., a
hypertetrahedron in $S-1$ dimensions.

If $\lambda $ is an eigenvalue of $\mathbb{K}$, so is $-\lambda $, since $%
\mathbb{K}$ is antisymmetric. Thus, if $S$ is odd, $\mathbb{K}$ must have at
least one zero eigenvalue. In case there is only one, let us denote its
associated left and right eigenvectors by: 
\begin{equation}
\left\langle \zeta \right| \mathbb{K}=\mathbb{K}\left| \zeta \right\rangle
=0\,\,.  \label{K zeta}
\end{equation}
Note that, in general, a left eigenvector is not simply related to the
transpose of the right eigenvector. However, since $-\mathbb{K}\left| \zeta
\right\rangle =0=\left( -\mathbb{K}\left| \zeta \right\rangle \right)
^T=\left( \left| \zeta \right\rangle \right) ^T\mathbb{K}$, $\left\langle
\zeta \right| $ is the transpose of $\left| \zeta \right\rangle $. These two
eigenvectors play ``dual'' roles in the following sense. Obviously, $\left|
\zeta \right\rangle $ is a fixed point of eqn. (\ref{X-eqn}) and, if every $%
\zeta _m$ is \emph{non-negative}
\footnote{More precisely, we need 
$\zeta _n\zeta _m$ to be non-negative for all $n,m$.}, 
then 
\begin{equation}
X_m^{*}\equiv \zeta _m\left/ \sum_\ell \zeta _\ell \right.   \label{FP}
\end{equation}
represents a fixed population
\footnote{Strictly, fixed points with some components being zero are also
uninteresting, since an extinct speices will remain so for ever. For
simplicity, readers may restrict their attention to points within the 
$S-1 $ simplex.}. If some are negative, then the system has no fixed
populations (except for absorbing states). Meanwhile, we also have 
$\partial \left. \left\langle \zeta \right| U\right\rangle =0$, so that 
\begin{equation}
\mathcal{R}\equiv \prod_m\left( X_m\right) ^{\zeta _m}  \label{R}
\end{equation}
is a constant in the evolution. In the three cyclically competing species
case\cite{BRSF09}, the invariant $R\equiv A^{k_b}B^{k_c}C^{k_a}$ can be
readily related to $\mathcal{R}$. When all $\zeta _m$'s are positive, we see
that these invariant manifolds (i.e., $\mathcal{R}$'s given by various
initial values of $X_m$) are closed $S-2$ dimensional subspaces within our $%
S-1$ simplex -- e.g., the closed loops in \cite{BRSF09}. For $S>3$, we can
be certain of the system evolving along orbits that lie in such a manifold,
but we have not shown whether these orbits are open or closed. In either
case, the long time average of $X_m\left( t\right) $, defined by 
\[
\stackrel{\circ }{\,X_m} \,\,\equiv \mathop{\lim}_{T\rightarrow \infty }
\frac 1T \int_0^TX_m\left( t\right) dt\,\,,
\]
is precisely $X_m^{*}$. The proof is simple: Integrate eqn. (\ref{U eqn}),
recognize the finiteness of $\ln X_m$, and invoke the uniqueness of $\left|
\zeta \right\rangle $. Note that $\mathcal{R}$ does change in a stochastic
model, as these orbits and the fixed point are \emph{neutrally stable} \cite
{DF10}. On the other hand, if some $\zeta _m$'s are negative, $\mathcal{R}$
is still meaningful and typically determine how pairs of species must vanish
in the long time limit. A good example is the three species case with say, $A
$ preying on $C$ with rate $k_c$ (instead of being consumed by $C$). Then, $C
$ must decrease monotonically. The (related) invariant is $%
R=A^{k_b}B^{-k_c}C^{k_a}$, so that $B$ and $C$ must vanish together, with $%
C\rightarrow \left[ RB^{k_c}\right] ^{1/k_a}$, leaving $A$ as the sole
survivor. Finally, if the spectrum of $\mathbb{K}$ contains $\nu >1$ zeros,
then there will be $\nu $ invariants and, ``dual'' to those, a $\nu -1$
dimension subspace of fixed points. Many of the conclusions can be readily
generalized. It would be interesting to explore further details of these
physical subspaces, especially if some of these eigenvectors have negative
components.

The case of even $S$ displays even richer behavior. First, it is possible
that none of the eigenvalues of $\mathbb{K}$ vanishes. In that case, it 
has a unique inverse, so that (\ref{U eqn}) can be written as 
\begin{equation}
\partial \mathbb{K}^{-1}\left| U\right\rangle =\left| X\right\rangle
\label{KU eqn}
\end{equation}
so that $\partial \left\langle 1\right| \mathbb{K}^{-1}\left| U\right\rangle =1$%
. Defining 
\begin{equation}
\eta _m\equiv \sum_\ell \left( K^{-1}\right) _{\ell m}  \label{eta}
\end{equation}
we see that the collective variable 
\begin{equation}
\mathcal{Q}\equiv \prod_m\left( X_m\right) ^{\eta _m}  \label{Q}
\end{equation}
evolves trivially: 
\begin{equation}
\mathcal{Q}\left( t\right) =\mathcal{Q}\left( 0\right) e^t  \label{Qexp}
\end{equation}
At first glance, this mathematical expression appears very ``rigid,'' as it
predicts an exponentially \emph{increasing} $\mathcal{Q}$ under all 
cirmstances. With a little reflection, 
a more ``flexible'' physical interpretation emerges: Whether a
species wins or loses is associated with the signs of the $\eta $'s.
Specifically, at least one of the species associated with \emph{negative} $%
\eta $'s must vanish at late times.

Let us turn to some simple examples to illustrate these findings. In the
most trivial case of $S=2$, let $x_1$ consume $x_2$ with rate $k>0$. Thus, $%
X_2$ must vanish at large $t$, consistent with $\mathcal{Q}$ being $\left(
X_1/X_2\right) ^{1/k}$. Indeed, the full solution is given simply by $%
X_1/X_2=e^{kt}\left. X_1/X_2\right| _{t=0}$. Since $X_1+X_2=1$, the explicit
solution follows from algebra. A particularly appealing form exploits the
symmetry $X_1\Leftrightarrow X_2\oplus t\Leftrightarrow -t$ and displays the
symmetric sigmoid associated with the evolution from $X_2\left( -\infty
\right) =1$ to $X_1\left( \infty \right) =1$. Defining 
\begin{equation}
\sigma \left( t\right) \equiv \tanh \frac{kt}2  \label{sigma}
\end{equation}
we easily find $X_1\left( t\right) -X_2\left( t\right) =\sigma \left(
t-t_0\right) $, with $t_0$ given by the initial condition: $\left(
2/k\right) \tanh ^{-1}\left[ X_2\left( 0\right) -X_1\left( 0\right) \right] $%
. Thus, the explicit solution is 
\begin{equation}
X_{1,2}\left( t\right) =\frac{1\pm \sigma \left( t-t_0\right) }2\,\,.
\label{x1}
\end{equation}

At the next level, $S=4$, we have the special case of four cyclically
competing species\cite{4SS}, where the variable $Q\equiv
A^{k_b+k_c}B^{-k_c-k_d}C^{k_d+k_a}D^{-k_a-k_b}$ can be readily related to $%
\mathcal{Q}$. Further, this system is an excellent example of a large class
of even $S$ systems, namely, ones with two ``opposing teams.'' Specifically,
suppose each ``team'' consists of $s=S/2$ species -- $\left\{ y_j\right\}
,\left\{ z_j\right\} \,;\,j=1,...,s$ -- with the understanding that species
on the same team \emph{do not} interact. Then, all the interactions can be
specified by the $s^2$ probabilities associated with encounters between $y_i$
and $z_j$. Note that the outcome can be either $2y_i$ or $2z_j$, depending
on which player is stronger. Note also that the absorbing states of the
stochastic model form two $s-1$ simplices. To proceed, we let $\left(
Y_i,Z_j\right) $ denote the fractions of the species $\left( y_j,z_j\right) $
and arrange the column vector according to 
\begin{equation}
\left| X\right\rangle \equiv \left( 
\begin{array}{c}
\left| Y\right\rangle  \\ 
\left| Z\right\rangle 
\end{array}
\right) \,\,.  \label{YZ}
\end{equation}
Then, $\mathbb{K}$ takes the form
\begin{equation}
\mathbb{K}=\left( 
\begin{array}{cc}
0 & \mathbb{M} \\ 
-\mathbb{M}^T & 0
\end{array}
\right) \,\,.  \label{K-M}
\end{equation}
Here $\mathbb{M}$ is a $s\times s$ matrix, the elements of which are associated
with the rate of $y_i$ consuming (or being consumed by) $z_j$. The
antisymmetry of $\mathbb{K}$ is ``built in'' and, unlike its spectrum, there
are no constraints on the spectrum of $\mathbb{M}$. Focusing first on
non-singular $\mathbb{M}$'s, we define 
\[
\Lambda \equiv \det \mathbb{M}\neq 0 
\]
(generalization of $\lambda $ in \cite{4SS}) and 
\begin{equation}
\mathbb{W}\equiv \Lambda \mathbb{M}^{-1}  \label{W}
\end{equation}
the elements of which are the appropriate co-factors of $\mathbb{M}$. 
These considerations allows us to find a non-trivial invariant manifold -- a
straight line joining the point
$\mathbb{W}^T\left| 1\right\rangle /\left\langle 1\right| \mathbb{W}^T\left|
1\right\rangle $ in the space of $\left| Y\right\rangle $ 
\footnote{Here, $\left| 1\right\rangle =\left\langle 1\right| ^T$. 
We also assume ${\mathbb{W}}^T\left| 1\right\rangle $ and 
$\mathbb{W}\left| 1\right\rangle $ have
positive elements. The cases where this is not true deserve further study.} 
with the point $%
\mathbb{W}\left| 1\right\rangle /\left\langle 1\right| \mathbb{W}\left|
1\right\rangle $ in the space of $\left| Z\right\rangle $. Indeed, the
evolution on this line is precisely the same as the $S=2s=2$ case, (\ref{x1}%
), i.e., 
\begin{equation}
\left| X\right\rangle =\frac 12\left( 
\begin{array}{c}
\left( 1+\sigma \right) \mathbb{W}^T\left| 1\right\rangle  \\ 
\left( 1-\sigma \right) \mathbb{W}\left| 1\right\rangle 
\end{array}
\right)   \label{two team}
\end{equation}
provided $k$ in (\ref{sigma}) is replaced by $\Lambda $ and an appropriate $%
t_0$ is taken. The proof is straightforward, since $\partial \ln \left[
\gamma f\left( t\right) \right] =\partial \ln f\left( t\right) $ for any
constant $\gamma $, so that 
\[
\partial \left| U\right\rangle =\left( 
\begin{array}{c}
\partial \ln \left( 1+\sigma \right) \left| 1\right\rangle  \\ 
\partial \ln \left( 1-\sigma \right) \left| 1\right\rangle 
\end{array}
\right) .
\]
Meanwhile, we have 
\[
\mathbb{K}\left| X\right\rangle =\frac \Lambda 2\left( 
\begin{array}{c}
\left( 1-\sigma \right) \left| 1\right\rangle  \\ 
-\left( 1+\sigma \right) \left| 1\right\rangle 
\end{array}
\right) .
\]
But $\partial \ln \left( 1\pm \sigma \right) =\pm \partial \sigma /\left(
1\pm \sigma \right) =\pm \Lambda \left( 1\mp \sigma \right) /2$, so that $%
\partial \left| U\right\rangle =\mathbb{K}\left| X\right\rangle $ is indeed
satisfied with $\sigma =\tanh \left( \Lambda t/2\right) $. The result (\ref
{two team}) is appealing, since $\Lambda =\det $ $\mathbb{M}$ is precisely $k$
in case $s=1$. In \cite{CDPZ-MFT}, the case of $s=2$ will be presented in
explicit detail. Note that we do not assume a sign for $\Lambda $ here. If $%
\Lambda >0$, team $\left\{ y_j\right\} $ will ``win,'' i.e., the system
ending at $\left| Z\right\rangle =0$ (and 
$\left| Y\right\rangle \propto \mathbb{W}^T\left| 1\right\rangle $). 
Similarly, if $\Lambda <0$, the system will end at 
$\left| Z\right\rangle \propto \mathbb{W}\left| 1\right\rangle $. Apart
from this special line, we can draw similar conclusions based simply on the
sign of $\Lambda $, namely, which team ``wins'' and which ``loses.'' We
should caution the reader that this way of displaying extinction is
restricted to systems with two teams, while eqns. (\ref{Q},\ref{Qexp}) are
valid in general.

Next, we turn to systems with singular $\mathbb{K}$'s, i.e., neutrally stable
ones. The null space of $\mathbb{K}$ must be even dimensional, a number we
denote by $2\mu $ (integer $\mu $). As in the odd $S$ case, the associated
left- and right-eigenvectors play ``dual'' roles in that they can be used to
construct $2\mu $ invariants as well as a $2\mu -1$ dimensional subspace of
fixed points (which may or may not be physical). A good illustration is the
special case of $S=2s$ \emph{cyclically competing} species, for which a
straightforward (though tedious) proof leads to $\mu $ being unity always 
\cite{CDPZ-MFT}. More generally, for systems with two opposing teams, we can
exploit (\ref{K-M}) and show explicit expressions for the subspace of fixed
points and invariants. Clearly, the methods for odd $S$ applies here: Define
the zero eigenvectors of $\mathbb{M}$ by 
\begin{equation}
\left\langle \tilde{\zeta}_\alpha \right| \mathbb{M}=\mathbb{M}\left| \zeta
_\alpha \right\rangle =0\,\,.  \label{M zeta}
\end{equation}
with $\alpha =1,...,\mu $. Note that, unlike $\mathbb{K}$, $\mathbb{M}$ is not
antisymmetric (or symmetric) in general, so that $\mathbb{M}^T$ is not
simply related to $\mathbb{M}$. Thus, 
$\left| \tilde{\zeta}_\alpha \right\rangle$ (i.e., the tranpose of 
$\left\langle \tilde{\zeta}_\alpha \right|$) 
and $\left| \zeta _\alpha \right\rangle $ are linearly independent,
typically. The subspace of fixed points is spanned by the $2\mu $ vectors 
\begin{equation}
\left( 
\begin{array}{c}
0 \\ 
\left| \zeta _\alpha \right\rangle
\end{array}
\right) ,\left( 
\begin{array}{c}
\left| \tilde{\zeta}_\alpha \right\rangle \\ 
0
\end{array}
\right)  \label{zeta}
\end{equation}
In case all the elements of these vectors are non-negative, this subspace
will have a non-trivial intersection with the physical region: $X_m\geq 0$
and $\sum_mX_m=1$. Then, the subspace of physical fixed points is $2\mu -1$
dimensional. In all cases, the $2\mu $ invariants defining orbits of neutral
stability can be written: 
\begin{equation}
\mathcal{R}_\alpha \equiv \prod_{j=1}^s\left( Z_j\right) ^{\left( \zeta
_\alpha \right) _j};\quad \tilde{\mathcal{R}}_\alpha \equiv
\prod_{j=1}^s\left( Y_j\right) ^{\left( \tilde{\zeta}_\alpha \right) _j}
\label{R's}
\end{equation}
showing again the ``dual'' role played by the zero-eigenvectors.

So far, we considered systems with non-singular and singular $\mathbb{K}$'s
separately. A unified approach can be formulated, taking advantage of the
cofactors, $\mathbb{W}$, being well defined even if $\Lambda =0$ and $\mathbb{M}%
^{-1}$ does not exist. Starting with eqns. (\ref{U eqn},\ref{U def}), we
multiply by

\begin{equation}
\left( 
\begin{array}{cc}
0 & -\mathbb{W}^T \\ 
\mathbb{W} & 0
\end{array}
\right) \,\,.  \label{W matrix}
\end{equation}
The right side is simply $\Lambda \left| X\right\rangle $ by virtue of eqn.
(\ref{W}), whether $\Lambda =0$ or not. Taking the inner product with 
$\left\langle 1\right| $, the right side reduces to $\Lambda $. Turning to
the left side, let us denote the elements of $\left\langle 1\right| \mathbb{W}$
and $\left\langle 1\right| \mathbb{W}^T$ by $\xi _j$ and $\chi _j$
respectively. These allow us to define two other collective variables 
\begin{equation}
F\equiv \prod_j\left( Y_j\right) ^{\xi _j};\quad G\equiv \prod_j\left(
Z_j\right) ^{\chi _j}  \label{FG}
\end{equation}
From these, we form 
\begin{equation}
Q\left( t\right) \equiv \frac FG=Q\left( 0\right) e^{\Lambda t}\,\,.
\label{capQ}
\end{equation}
which is readily seen as $\mathcal{Q}^\Lambda $ but is more ``flexible'' --
in that $Q$ can increase or decrease with $t$. Similarly, it is
straightforward to show that, as $\Lambda \rightarrow 0$, $F$ and $G$
reduces to $\mathcal{R}$ and $\tilde{\mathcal{R}}$ in (\ref{R's}). In this
sense, the variables (\ref{FG},\ref{capQ}) are superior to 
(\ref{R's},\ref{Q}).

\emph{Concluding remarks.} In this brief note, we report findings of general
properties associated with the rates equations for a system of $S$ species
that compete pairwise. In some cases (all odd $S$ and some even $S$),
subspaces of fixed points appear to be ``dual'' to invariant manifolds. In
the absence of this behavior, we find a collective variable which evolves
exponentially. We conclude with a few remarks on possible future research,
apart from the issues already raised above.

In a standard Lotka-Volterra model of predator-prey interactions,
individuals can be born and can die without the interaction with those of
another species, as in (\ref{bd}). Thus, an equation like (\ref{X-eqn})
should contain linear terms 
\begin{equation}
\partial N_m=\gamma _mN_m+N_m\sum_\ell k_{m\ell }N_\ell  \label{LV-eqn}
\end{equation}
where $\gamma _m$ is $b_m$ or $-d_m$. Of course, we can re-analyze our
problem with these additions. However, it is possible to regard this problem
as a special limit of an $S+1$ species system. First, note that $\gamma _m$
are generally chosen to be similar to $k_{m\ell }$, so that the typical $N_m$%
's reflect the numbers in nature (thousand, million, etc.). For example, in
the $S=2$ case, the typical levels of the predator and prey are $d/k$ and $%
b/k$, respectively. Next, introduce $N_0$ individuals of species $x_0$, let
them interact with $x_{m\neq 0}$ according to 
\begin{equation}
x_m+x_0\stackrel{b_m}{\longrightarrow }2x_m\,;\quad 
x_m+x_0\stackrel{d_m}{\longrightarrow }2x_0  \label{0mpair}
\end{equation}
and consider the limit of $N,N_0\rightarrow \infty $. Thus, $X_0=O\left(
1\right) $ and $X_{m\neq 0}=O\left( 1/N\right) $. Now, consider a
generalized eqn. (\ref{X-eqn}): 
\begin{equation}
\partial X_m=X_m\sum_\ell L_{m\ell }X_\ell  \label{L-eqn}
\end{equation}
with $m,\ell =0,1,...,S$ and 
\begin{eqnarray*}
L_{m\ell } &=&Nk_{m\ell }\quad m,\ell \neq 0 \\
L_{m0} &=&-L_{0m}=\gamma _m
\end{eqnarray*}
Keeping lowest order terms in each equation, we find a consistent limit for (%
\ref{LV-eqn}), i.e., 
\begin{equation}
\partial X_0=0+O(1/N)  \label{0}
\end{equation}
(which admits $X_0=1+O(1/N)$) and, for $m\neq 0$, 
\begin{equation}
\partial X_m=\gamma _mX_m+X_m\sum_\ell k_{m\ell }N_\ell +O(1/N^2)  \label{m}
\end{equation}
In other words, the evolution of an ordinary Lotka-Volterra system takes
place in a tiny ``corner'' of the enlarged configuration space. In
particular, the familiar fixed point and closed orbits in the 2 species
Lotka-Volterra system ($d/k,b/k$) are just the ones in the case of 3
cyclically competing species.

Since all realistic evolution is stochastic, this study should be extended,
along the lines in \cite{DF10}, for example. The invariants here are clearly
related to slow variables and identifying the fast variables will be
helpful. For systems with two teams, expression (\ref{K-M}) behooves us to
explore a possible simplectic structure in our dynamics, even though we must
deal with difficulties associated with non-linearities. In several special
cases, the implications of Nambu dynamics for this kind of evolution was
studied \cite{FKB96}. Since that approach requires $S-1$ invariants
(``Hamiltonians''), we can easily see if it can be exploited for any given
set $\left\{ p_m^\ell \right\} $, by checking whether the spectrum of $\mathbb{K%
}$ has $S-2$ zero's. Beyond such possibilities, an entirely new vista of
systems with competing species awaits us, e.g., ones with non-trivial
spatial structures, ones on complex networks, and ones with inhomogeneous
rates. Clearly, to reflect the enormous range of phenomena in nature, we can
create many more models and can expect many more interesting results.

\emph{Acknowledgments.} I thank Erwin Frey, Kirone Mallick, Sid Redner, 
Eric Sharpe, Zoltan Toroczkai, Michel Pleimling, and other members of the 
statistical physics group at Virginia Tech for illuminating discussions. 
This work is supported in part by the US National Science Foundation 
through Grants DMR-0705152 and DMR-1005417. \newline \newline 

\emph{References.} \newline

\end{document}